\documentstyle[12pt]{article}
\def\bc{\begin{center}}      \def\ec{\end{center}}
\def\i{\indent}              \def\nnb{\nonumber}
\def\vs{\vspace}                
\def\dg{\dagger}          \def\la{\langle}       \def\ra{\rangle}
\def\l{\left}                \def\r{\right}
\def\be{\begin{equation}}    \def\ee{\end{equation}}
\def\bear{\begin{eqnarray}}  \def\eear{\end{eqnarray}}
\def\lb{\label}              \def\ci{\cite}
\def\pr{\prime}           \def\tl{\tilde}      \def\lam{\lambda}
\def\Lam{\Lambda}         \def\alf{\alpha}      \def\sig{\sigma}
\def\bt{\begin{tabular}}      \def\et{\end{tabular}}

\begin{document}
\begin{flushright} preprint INRNE-TH-97/1 (quant-ph/9701018)\\
			   (June, 1997: minor changes in text and references)\\
				   Accepted in Journal of Physics A
\end{flushright}					    	
\vs{5mm}

\bc
		{\large \bf ROBERTSON INTELLIGENT STATES}\\[2mm]
                   {\bf D.A. Trifonov}\\
                Institute for Nuclear Research\\
                  72 Tzarigradsko chauss\'ee\\
                    1784 Sofia, Bulgaria
 \ec
\vs{5mm}
\abstract
	Diagonalization of uncertainty matrix and minimization of Robertson
inequality for $n$ observables are considered. It is proved that for even
$n$ this relation is minimized in states which are eigenstates of $n/2$
independent complex linear combinations of the observables. In case of
canonical observables this eigenvalue condition is also necessary. Such
minimizing states are called Robertson intelligent states (RIS).
The group related coherent states (CS) with maximal symmetry (for
semisimple Lie groups) are particular case of RIS for the quadratures of
Weyl generators.  Explicit constructions of RIS are considered for
operators of $su(1,1)$, $su(2)$, $h_N$ and $sp(N,R)$ algebras. Unlike the
group related CS, RIS can exhibit strong squeezing of group generators.
Multimode squared amplitude squeezed states are naturally introduced as
$sp(N,R)$ RIS.  It is shown that the uncertainty matrices for quadratures
of $q$-deformed boson operators $a_{q,j}$ ($q>0$) and of any $k$ power of
$a_j=a_{1,j}$ are positive definite and can be diagonalized by symplectic
linear transformations.

\vs{5mm}
\normalsize
\baselineskip 18pt

\section{Introduction}

Canonical coherent states (CS) \ci{GlKl,KlSk} in quantum optics and
quantum mechanics can be defined in three equivalent ways:
1) as eigenstates of nonhermitean boson (photon) destruction operator $a$;
2) as orbit of the oscillator ground state $| 0\ra$ under the action of
   unitary displacement operator $D(\alpha)$;
3) as states which minimize  the Heisenberg relation for canonical
observables $q=(a+a^\dg)/\sqrt{2}$ and $p = -i(a-a^\dg)/\sqrt{2}$ with
equal uncertainties.  Correspondingly there are three ways of
generalizations of canonical CS \ci{Feng}.  As most general up to now is
considered the second one, which consists in construction orbits of a
reference vectors $|\psi_0\ra$ under the action of unitary operators of
irreducible representations of a given Lie group \ci{KlSk,Feng}
($D(\alpha)$ realize ray representation of Heisenberg--Weyl group $H_1$).
These generalized CS are known (and should be called here) as group
related CS \ci{KlSk}.

The main aim of the present paper is to consider the third way of
generalization (the intelligence way) to the case of $n$ observables and
its relationship to the first two ones. The idea is to look for a
generalized uncertainty relation for $n$ observables $X_\mu$,
$\mu=1,2,\ldots,n$, which minimization would yield a continuous family of
states such that in cases of $X_\mu$ being generators of a Lie group it
would include the corresponding group related CS.

It turned out that the required generic uncertainty relation (UR) for $n$
observables is that of Robertson \ci{Ro}, eq.  (\ref{RUR1}), (see also the
review \ci{183} on generalized uncertainty relations). Here we show that
it is minimized in the eigenstates of $n/2$ (for even $n$) independent
complex linear combinations of $X_\mu$ or (for any $n$) of at least one
real combination. When $X_\mu$ are quadrature components of
Weyl generators of a semisimple Lie group \ci{BaRo} these minimizing states
contain as subset the corresponding group related CS with symmetry
\ci{KlSk,Feng}. Thus it is Robertson relation that naturally connects the
above three ways of generalization of CS on the level of $n$ observables.
In case of $N$ mode electromagnetic field we get that Robertson UR (RUR) is
minimized if and only if the state is an eigenstate of $N$ new destruction
operators $a^\pr_j = u_{jk}a_k+v_{jk}a^\dg_k$.  For brevity states which
minimize some uncertainty relation should be called here {\it intelligent
states} (IS) (the term IS is introduced in \ci{Ar} on the example of spin
states which minimize Heisenberg UR). The term {\it correlated}
\ci{DKM} is reserved for states with nonvanishing covariances
(correlations).

The first step on the first way of CS generalization was made in papers
\ci{MMT69,MMTStoLu,MMTHo} where eigenstates of complex
combinations of $a_j$ and
$a^\dg_j$, $j=1,2\ldots,N$, were constructed and discussed ($N=2$
in \ci{MMT69}, $N=1$ in \ci{MMTStoLu}, any $N$ in \ci{MMTHo}). Later \ci{DKM}
it was shown that eigenstates of $ua+va^\dg$ minimize the
Schr\"odinger UR (SUR) \ci{Sch} for $q$ and $p$ (eq. (\ref{SUR})), the
minimizing states being called {\it correlated} CS. Those  CS in fact are
the same \ci{Tr93} as canonical squeezed states in quantum optics
\ci{LoKn}. In \ci{Tr94} it was proved that SUR for any
two observables $X$ and $Y$ is minimized in eigenstates of their complex
combination $\lam X+iY$, (equivalently of $uA+vA^\dg$,
$A=(X+iY)/\sqrt{2}$, $\lam$ and $u,\,v$ are complex numbers).  Eigenstates
of $uA+vA^\dg$ can exhibit strong squeezing in $X$ and $Y$. Schr\"odinger
IS (SIS) for the generators $K_1,\,K_2$ of $SU(1,1)$ were constructed in
\ci{Tr94} and shown to combine the Barut--Girardello CS \ci{BaGi} and
$SU(1,1)$ group related CS with symmetry \ci{Feng}. The full sets of even
and odd SIS for quadratures of squared boson destruction operator $a^2$
were constructed in the second paper of ref. \ci{SCB-Tr95} (see also
\ci{Tr96,Br96}).  Eigenstates of $uA+vA^{\dg}$ with real $u$ and $v$ are
noncorrelated SIS, that is Heisenberg IS, and the cases of $X,\,Y$ being
quadratures of $aa$ or of the product $ab$ of two annihilation operators
were considered in papers \ci{BeHiYu}.

Another purpose of the paper is to consider the diagonalization problem of
uncertainty matrix, denoted here by $\sigma$. This matrix is of direct
physical significance since its elements are dispersions (variances) and
correlations (covariances) of observables. $\sigma$ is important also in
quantum state geometry \ci{ProVal}. In case of canonical operators $q_j$,
$p_k$ diagonalization of $\sigma$ was considered in \ci{SCB-Tr95}.

The paper is organized as follows. In section 2 we briefly review
Robertson relations for the uncertainty matrix $\sigma$ for $n$
observables $X_\mu$. In section 3 we consider the diagonalization of
$\sigma$ by means of linear transformations of $X_\mu$. We note that in
any state $\sigma$ can be diagonalized by means of orthogonal
transformation. From this it follows that the spin component correlations
can be eliminated by coordinate rotation. When the uncertainty matrix is
positive definite (as is the case of $2N$ quadratures of $k$ power of
boson/photon annihilation operators $a_j$ and the case of quadratures of
$q$-deformed boson operators $a_{j,q}$ for $q>0$) it can be diagonalized
also by means of symplectic transformation.  New family of trace class UR
(\ref{newur2}) is established for positive definite dispersion matrices.

In section 4 we study the minimization of $n$ dimensional RUR.  In section
5 explicit examples of RIS are considered, the $su(1,1)$ and $su(2)$ RIS
being discussed in greater detail. RIS for generators of $SU(1,1)$ in
quadratic bosonic representation can exhibit linear and quadratic
amplitude squeezing (even simultaneously - joint squeezing of two
noncommuting observables).

\section{Robertson uncertainty inequalities}

For $n$ observables (hermitean operators) $X_\mu$ Robertson \ci{Ro} (see
also review paper \ci{183}) established the following two uncertainty
relations for the dispersion matrix $\sigma$,
\be \lb{RUR1}   
\det \sigma \geq \det C,
\ee
\be \lb{RUR2}   
\sigma_{11}\sigma_{22}...\sigma_{nn}\geq \det \sigma,
\ee
where $\sigma_{\mu\nu} = \la X_\mu X_\nu+X_\nu X\mu\ra/2 -\la X_\mu\ra\la
X_\nu\ra$ and $C$ is the antisymmetric matrix of mean commutators,
$C_{\mu\nu} = -(i/2)\la[X_\mu,X_\nu]\ra$. Here $\la X\ra$ is the mean
value of $X$ in quantum state $\rho$, which is generally mixed state. For
$n=2$ inequality (\ref{RUR1}) coincides with SUR ($X_1 = X,\,\,X_2=Y$)
\be\lb{SUR}  
\Delta^2X\Delta^2Y - \sig^2XY \geq \frac{1}{4}|\la[X,Y]|^2,
\ee
which in turn is reduced to the Heisenberg UR for $X$ and $Y$
when the covariance $\sigma_{XY} = \la XY+YX\ra/2-\la X\ra\la
Y\ra$ is vanishing ($\Delta^2X\equiv\sig_{XX}$, $\Delta^2Y\equiv\sig_{XY}$).

Combining (\ref{RUR1}) and (\ref{RUR2}) one gets
\be\label{HUR}   
\sigma_{11}\sigma_{22}...\sigma_{nn}\geq \det C,
\ee
which can be treated as direct extension of Heisenberg UR to the
case of $n$ operators.

The uncertainty matrix (dispersion or correlation matrix) $\sigma =
\sigma(\vec{X},\rho)$, where $\vec{X} = (X_1,X_2,\ldots, X_n)$, is
symmetric by construction.  From Robertson inequalities
(\ref{RUR1}), (\ref{RUR2}) one can deduce that its determinant is always
nonnegative.  Indeed, the matrix of mean commutators is antisymmetric and
the determinant of antisymmetric matrix is nonnegative \ci{Ga}. Thereby in
any state $\rho$ we have $\det C \geq 0$. $\det C$ vanishes identically if
the number of operators $n$ is odd.

Diagonal elements of $\sigma$ are the variances of $X_\mu$. The problem of
reducing (squeezing) of variances of quantum observables is of importance
in physics (in quantum optics \ci{LoKn}) of precise measurements and
telecommunications. The nondiagonal elements are the covariances of $X_\mu$
and $X_\nu$ and describe $X_\mu$-$X_\nu$ correlations. The uncertainty
matrix in pure state $|\psi_0\ra$ can be used as a metric tensor in the
manifold of generalized Glauber CS $D(\alpha)|\psi_0\ra$ \ci{ProVal}. In
view of these dynamical and geometrical properties of $\sigma$ it is
desirable to study the problem of its diagonalization (which is equivalent
to the problem of minimization of the second Robertson relation
(\ref{RUR2})). Diagonalization of $\sigma$ in the case of canonical
observables $p_j=X_j$, $q_j=X_{N+j}$, $j=1,2,\ldots,N$, was recently
considered in \ci{SCB-Tr95}: in any state it can be diagonalized by means
of linear canonical transformations. In the next section we consider this
problem in more general cases.  The minimization of (\ref{RUR1}) for two
observables $X_1$ and $X_2$ (i.e. of SUR (\ref{SUR})) has been shown
\ci{Tr94}  to  occur  in  the eigenstates of their complex (in
particular real) linear combinations only. In section  4  we extend this
result to arbitrary $n$.

\section{Diagonalization of uncertainty matrix of $\qquad\qquad\qquad$
$n$ observables}

In this section we consider the diagonalization of the uncertainty matrix
$\sigma(\vec{X},\rho)$ by means of linear
transformations of $n$ operators $X_\mu$ (summation over repeated
indices),
\be \lb{LT1}  
X_\mu \rightarrow X^\pr_\mu = \lam_{\mu\nu} X_\nu,
\ee
where $\lam_{\mu\nu}$ are real numbers (in order $X^\pr_\mu$ to be again
hermitean operators).

We first note the transformation property of $\sigma$ under transformation
(\ref{LT1}). Defining the new matrix $\sigma^\pr$ as $\sigma^\pr =
\sigma(\vec{X^\pr},\rho)$ we easily get
\be \lb{sig1}  
\sigma^\pr = \Lam \sigma\Lam^{T},
\ee
where we introduced $n$ vector $\vec{X}^\pr = (X^\pr_1,\ldots,X^\pr_n)$
and $n\times n$ matrix $\Lam = \{\lam_{\mu\nu}\}$, its
transposed being denoted as $\Lam^{T}$.  Thus the two dispersion matrices
are congruent via the transformation matrix $\Lam$. We suppose that
transformation (\ref{LT1}) is invertable and set $\det\Lam =1$. In matrix
form eq. (\ref{LT1}) is rewritten as $\vec{X}^\pr = \Lam \vec{X}$.

We note several general properties of $\sigma$, some of which being immediate
consequences of its symmetricity and the transformation law (\ref{sig1}).
First we note the invariant quantities: a) $\det\sigma = \det\sig^\pr$ for
any $\Lam \in SL(n,C)$; b) Tr$\sigma^k={\rm Tr}\sig^\pr{}^k$,
$k=1,2,\ldots$, for orthogonal $\Lam$; c) Tr$(\sig J)^k={\rm Tr}(\sig^\pr
J)^k$ for symplectic transformations ($n=2N$),
\be\lb{J}     
\Lam J\Lam^T = J,\quad J=\l(\bt{ll}
$0$&$1_N$\\$-1_N$&$0$\et\r).
\ee
The last two invariants are particular cases of quite general relations
Tr$(\sig^\pr g)^k = {\rm Tr}(\sig g)^k$ which hold for $\Lam$ satisfying
$\Lam^Tg\Lam = g$ with any fixed matrix $g$ (in the above $g=1$ and
$g=J$).

Next we note that $\sig$ (being symmetric) can be always diagonalized by
means of orthogonal $\Lam$ ($\Lam\Lam^{ T}=1$) \ci{Ga} in any state, i.e.
$\sigma^\pr$ is diagonal for some orthogonal $\Lam$.  In case of spin (or
angular momentum) operators we get from this property that spin component
correlations can be considered as pure coordinate effects.  An other
general property of $\sigma$ is its nonnegativity, $\sigma \geq 0$.  To
prove this last property we diagonalize $\sigma$ by means of orthogonal
matrix $\Lam$. The new operators $X_\mu^\pr$, eq.  (\ref{LT1}), are again
hermitean and therefor all the diagonal elements of the matrix
$\sigma^\pr$ are nonnegative.  Therefor $\sigma \geq 0$ in any state
$\rho$.

Further properties of the uncertainty matrix can be established when the
set of operators $X_\mu$ possess some additional properties.  For example
if $\sigma$ is positive definite, $\sigma>0$, then it can be diagonalized
by means of {\it symplectic} $\Lam$ \ci{CoBoGo}.  Therefor it is important
to know when the uncertainty matrix is strictly positive. The value of
$\det C \geq 0$ turned out to play important role. Note that
$\det\sig>0$ stems from $\sig >0$ and $\det\sig=0$ means that $\sig$ is
not strictly positive. \\[-1mm]

{\large Proposition 1.} {\it $\det\sig(\vec{X},\rho)=0$ in pure
states $\rho = |\psi\ra\la\psi|$ if and only if
$|\psi\ra$ is an eigenstate of a real combination $\lam_\nu X_\nu$
 of $X_\nu$}.

{\it Proof}. a) Necessity. Let $\det\sig(\vec{X},\rho)=0$. Then orthogonal
$\Lam$ exists such that $\sig^\pr$ is diagonal. We have $0=\det\sig =
\det\sig^\pr = \sig^\pr_{11} \sig^\pr_{22} ...\sig^\pr_{nn}$, wherefrom at
least for one $\nu$ one has $\sig^\pr_{\nu\nu}=0$. The latter is possible in
pure states $\rho = |\psi\ra\la\psi|$ if and only if $X^\pr_\nu|\psi\ra =
x^\pr_\nu|\psi\ra$.  b) Sufficiency.  Let $(\vec{\lam}\vec{X})|\psi\ra =
x^\pr|\psi\ra$, $\vec{\lam}\vec{X} \equiv \lam_\nu X_\nu$. Then we can
always construct nondegenerate matrix $\Lam$ with first row
$(\lam_1,\lam_2,\ldots,\lam_n)$ and consider the uncertainty matrix
$\sig^\pr = \sig(\Lam\vec{X};\psi)$.  This $\sig^\pr$ is with vanishing
determinant since the first column of it is zero (as a consequence of
$(\vec{\lam}\vec{X})|\psi\ra = x^\pr|\psi\ra$). But $0=\det\sig^\pr =
(\det\Lam)^2\det\sig$, therefor $\det\sig=0$.

In view of this proposition and eq. (\ref{RUR1}) one has the \\[-1mm]

{\large Corollary 1}: {\it If $\det C(\vec{X},\psi) > 0$ then $|\psi\ra$
can't be normalizable eigenstate of any real combination $\lam_\nu
X_\nu$.}

If $\det C(\vec{X};\psi) > 0$ in any state then neither $X_\mu$ nor any
real combination $\lam_\nu X_\nu$ can be diagonalized in Hilbert space of
states $\cal{H}$, that is the spectrum of $X_\mu$ and $\lam_\nu X_\nu$ are
continuous.  Here is a class of $2N$ operators for
which  $\det C > 0$ and therefor $\sig$ is positive in any state.\\[-1mm]

{\large Proposition 2}. {\it If $X_\mu$, $\mu=1,2,\ldots,2N$ obey the
commutation relations
\be \lb{cr1}  
[X_j,X_{n+k}]=\delta_{jk}[X_j,X_{n+j}],\quad [X_j,X_{k}]=0=[X_{n+j},X_{n+k}],
\ee
where $-i[X_j,X_{n+j}]$ are positive definite operators, then $\det
C(\vec{X};\rho)>0$ and
the uncertainty matrix $\sig(\vec{X};\rho)$ is positive definite}.

{\it Proof}. By direct calculations we get
\be\label{detC}   
\det C = \l(\frac{1}{2}\r)^{2N}
\la(-i)[X_1,Y_1]\ra^2.\la(-i)[X_2,Y_2]\ra^2...\la(-i)[X_N,Y_N]\ra^2,
\ee
where $Y_j \equiv X_{N+j}$, $j = 1,2,\ldots,N$. Since every factor in
(\ref{detC}) is positive one has $\det C >0$.  From  corollary 1 and
the diagonalization of $\sig$ by orthogonal $\Lam$ we derive that $\det C >
0$ is a sufficient condition for $\sig$ to be positive definite.
End of proof.

We can point out a family of boson system (e.g. $N$ mode electromagnetic
field) observables which obey the commutation relations (\ref{cr1}). Those
are the quadrature components of power $k$ of photon (boson) destruction
operators $a_j$, defined here as
\be\label{op1}   
X^{(k)}_j= \frac{1}{\sqrt{2k}}(a_j^k+a_j^{k\dg}), \quad
X^{(k)}_{N+j}=\frac{-i}{\sqrt{2k}}(a_j^k-a_j^{k\dg})\equiv Y_j.
\ee
The relations (\ref{cr1}) and the positivity of $-i[X^{(k)}_j,Y^{(k)}_j] =
(1/k)[a^k,a^{k\dg}]$ can be checked by direct calculations.  As a result
the quadrature components of $a^k$ are continuous observables, their
uncertainty matrix is positive definite and can be diagonalized by means
of symplectic $\Lam$.  For $k=1$ operators (\ref{op1}) are the canonical
pairs $q_j$, $p_j$, therefor their uncertainty matrix can be diagonalized
by means of linear canonical transformations, corresponding to symplectic
$\Lam$. The  procedure for diagonalization of positive definite matrix by
means of symplectic $\Lam$ is described in \ci{CoBoGo} and in
the first paper of ref. \ci{SCB-Tr95}. Canonical transformations with time
dependent $\Lam(t)$ can be used to diagonalize any quadratic Hamiltonian.
For oscillator with varying mass and/or frequency this is done by
Seleznyova \ci{Se}.

Positive definite uncertainty matrices exist also in q-deformed boson
systems. $q$-deformed oscillator is introduced in \ci{MaBi}. The
deformed lowering and raising operators $a_q$ and $a^\dg_q$ obey the
commutation relation
\be\lb{qcr1}  
[a_q,a^\dg_q] = [N_q+1] - [N_q],\quad [N] \equiv
\frac{q^N-q^{-N}}{q-q^{-1}},
\ee
where $N_q$ is a number operator which eigenstates are $| n\ra_q =
([n]!)^{-1/2}a_q^{\dg n}| 0\ra_q$: $\quad$ $N_q| n\ra_q = n|n\ra_q,\quad
a_q| 0\ra_q =0$, $[n]! = [n][n-1]\ldots[1]$. At $q=1 \quad a_q,\,a^\dg_q$
coincide with ordinary boson operators $a,\,a^\dg$. Now we note that the
commutator $[a_q,a^\dg_q]$ is positive definite for $q>0$ as one can
easily verify, using (\ref{qcr1}).  From the commutation relations for $n$
$q$-deformed oscillators
\ci{Oh}
\bear\lb{qcr2}  
[a_{q,j},a_{q,k}] = 0, \quad
[a_{q,j},a^\dg_{q,k}] = \delta_{jk}[a_{q,j},a^\dg_{q,j}] \\
\, [N_{q,j},a_{q,k}] = - \delta_{jk}a_{q,k},\quad
[N_{q,j},a^\dg_{q,k}] = \delta_{jk}a^\dg_{q,j} \nnb
\eear
it follows that the set of quadrature components of $a_{q,j}$
obey the requirements of proposition 2 for $q>0$. Therefor the uncertainty
matrix $\sig(\vec{X}_q;\rho)$ is positive definite in any state for $q>0$.

For positive definite uncertainty matrix of $2N$ observables satisfying
(\ref{cr1}) one can establish a set of new uncertainty relations. In this
purpose consider the invariant quantities ${\rm Tr}(i\sig J)^{2k}$,
$k=1,2,\ldots,$. Let $\sig^\pr$ be diagonal matrix which is symplectically
congruent to $\sig$. Then we have
\be\lb{newur}   
 {\rm Tr}(i\sig J)^{2k}={\rm Tr}(i\sig^\pr J)^{2k}=2\sum_j^N
[\sig^\pr_{j,j}\sig^\pr_{N+j,N+j}]^k.
\ee
In view of $\sig > 0$ every term $\sig^\pr_{j,j}\sig^\pr_{N+j,N+j}$ in
(\ref{newur}) is nonvanishing and positive. We can apply the Heisenberg
relation for $\sig^\pr_{j,j}\sig^\pr_{N+j,N+j}$ and write the set of
inequalities
\be\lb{newur1} 
 {\rm Tr}(i\sig J)^{2k} \ge
2\sum_j^N\l|\frac{1}{2}\la[X^\pr_j,X^\pr_{N+j}]\ra\r|^k.
\ee
In the above $X^\pr_\mu = \Lam_{\mu\nu}(\rho)X_\nu$ and $\Lam(\rho)$ is
the diagonalizing symplectic matrix for the state $\rho$.  For every state
we can in principle find the minimal value $c^2_0(\rho)$ of the $N$ terms
$\l|\la[X^\pr_j,X^\pr_{N+j}]\ra\r|$ and therefor rewrite (\ref{newur1}) in
a more compact form
\be\lb{newur2}    
 {\rm Tr}(i\sig J)^{2k}\geq \frac{N}{2^{2k-1}}\,c_0^{2k},\quad k=1,2,\ldots.
\ee
In particular case of canonical variables $X_j=p_j$, $X_{N+j}=q_j$ in any
state the products $\sig^\pr_{j,j}\sig^\pr_{N+j,N+j}$ are greater or equal
to $1/4$ (this is the value of $\sig^\pr_{j,j}\sig^\pr_{N+j,N+j}$ in
Glauber CS for mode $j$, $\hbar=1$), that is $c^2_0 \geq 1$. Thus for
canonical variables  the above UR read
($\vec{Q}=(p_1,\ldots,p_N,q_1,\ldots,q_N$))
\be\lb{newur3}     
{\rm Tr}[i\sig(\vec{Q},\rho)J]^{2k}\geq \frac{N}{2^{2k-1}}.
\ee
The latter inequalities for the case of $\sig(\vec{Q},\rho)$ (apart from
the factor $i$) were recently obtained by Sudarshan, Chiu and Bhamathi
\ci{SCB-Tr95}. For $N=1$ and $k=1$ the inequality (\ref{newur3}) recovers
the Schr\"odinger relation (\ref{SUR}).

The above considered diagonalization of uncertainty matrix of $n$
hermitean operators by means of transformations of operators $X_\mu
\rightarrow X^\pr_\mu$ should be referred here as first kind
diagonalization. The state $\rho$ here is kept the same. This
diagonalization is always possible as we have shown. But it is of interest
also to know when $\sigma$ can be diagonalized by state transformation,
keeping observables the same. That is for given $X_\mu$ and state $\rho$
to find new state $\rho^\pr$ so that the new matrix $\sigma^{\pr\pr}
\equiv \sigma(\vec{X},\rho^\pr)$ be diagonal. We shall call this second
kind diagonalization.
Evidently both diagonalizations coincide (i.e. $\sigma^\pr =
\sigma^{\pr\pr}$) when the transformation (\ref{LT1}) is generated by some
unitary operator $U(\Lam)$,
\be \lb{LT2}   
X^\pr_\mu = \lam_{\mu\nu} X_\nu = U^\dg(\Lam)X_\mu U(\Lam).
\ee
Such is the case e.g. of uncertainty matrix $\sigma(\vec{Q},\rho)$ of
canonical operators $p_j\equiv Q_j$ and $q_j\equiv Q_{N+j}$ when the
diagonalizing $\Lam$ is symplectic. Then $U(\Lam)$ is a representation of
the group $Sp(N,R)$ \ci{BaRo} (more precisely of $Mp(N,R) =
\overline{Sp(N,R)}$) and thus any pure or mixed canonical correlated
states is  unitary
equivalent to noncorrelated state.  In case of $N=1$ we have an extra
diagonalizing property: in view of the fact that the squared boson
operators $a^2$, $a^{\dg 2}$, $a^\dg a$ close the $su(1,1)$ algebra, eq.
(\ref{sp1op}),  ($su(1,1)\sim sp(1,R)$) we get that in the one mode field
case the quadratic amplitude dispersion matrix is also diagonalizable by
unitary $Mp(1,R)$ state transformation.  The property (\ref{LT2}) occurs
also in the cases when $X_\mu$ close orthogonal algebra $so(n,R)$. Then
the diagonalizing orthogonal transformation (\ref{LT1}) is generated by
unitary $U(\Lam)\in SO(n,R)$.  On the example of $so(3,R)\sim su(2)$ this
means (recall that if $J_k$ are spin operators, $[J_k,J_j] = i\hbar
\epsilon_{kjl}J_l$, and $\Lam$ is orthogonal then $[J^\pr_k,J^\pr_j] =
i\hbar \epsilon_{kjl}J^\pr_l$) that spin component correlations
(covariances) in any state can always be eliminated by means of coordinate
rotation (first kind diagonalization) and by state transformation with
unitary operator $U(\Lam)$ (second kind diagonalization). In other words
spin component correlation is pure coordinate effect and any spin
correlated state is unitary equivalent to a noncorrelated one.

\section{Minimization of Robertson uncertainty $\qquad\qquad\qquad\qquad$
inequality $\det\sig \geq \det C$}

One general sufficient condition for minimization of Robertson inequality
(\ref{RUR1}) for arbitrary observables $X_\mu$, $\mu =1,2,\ldots,n$,
follows from the proposition 1: the equality in (\ref{RUR1}) holds in the
eigenstates of at least one of $X_\mu$ since in such case both matrices
$\sigma$ and $C$ have at least one vanishing column and then $\det\sig =
\det C = 0$.  In view of the fact that $\sigma$ can be always digonalized
by means of orthogonal $\Lam$ (second immediate property in section 2) the
minimization of both Robertson relations for any $n$ occurs also in the
eigenstates of some of $X^\pr_\mu =\lam_{\mu\nu} X_\nu$.

In case of odd $n$ the above sufficient condition for minimization of
(\ref{RUR1}) is also a necessary one: The inequality (\ref{RUR1}) is
minimized in a state $|\psi\ra$ if and only if $|\psi\ra$ is eigenstate of
a real combination $\lam_\nu X_\nu$ of observables $X_\nu$. The proof
follows from the proposition 1 and the property of determinant of
antisymmetric matrices of odd dimension: for odd $n$  $\det C$ of
antisymmetric matrix $C$ is vanishing identically in any state.

$\det C$ can be greater than $0$ for even $n$ only.  For even number of
operators $X_\mu$ we establish the following sufficient condition.
\\[-1mm]

{\large Proposition 3.} {\it The equality in the RUR
(\ref{RUR1}) for $2N$ hermitean operators $X_\mu$ holds in the eigenstates
$|\psi\ra$ of $N$ independent complex linear combinations of $X_\mu$ }.

{\it Proof}.
Let $X^\pr_\mu = \lam_{\mu\nu} X_\nu\equiv X^\pr_\mu(\Lam)$ be some linear
transformation
which preserves the hermiticity, i.e. $\lam_{\mu\nu}$ are real parameters.
We introduce $N$ nonhermitean operators $A_j = X_j+iX_{N+j}$ and construct
$N$ independent complex combinations of all $X_\nu$ in the form ,
\be\label{A-Apr}   
 A_j^\pr = X^\pr_j + iX^\pr_{N+j} = u_{jk}A_k+v_{jk}A^\dg_k,
\ee
where $u_{jk}$ and $v_{jk}$ are new complex parameters which are simply
expressed in terms of $\lam_{\mu\nu}$ ($j,k = 1,2,\ldots,N$). Let now
$|\psi\ra$ be eigenstate of all $A_j^\pr$,
\be\label{eve1}  
A_j^\pr|\psi\ra = z_j|\psi\ra, \quad j=1,2,\ldots,N,
\ee
$z_j$ being the eigenvalue. It is natural to denote the solutions of
(\ref{eve1}) as $|\vec{z},u,v\ra$ or equivalently as $|\vec{z},\Lam\ra$,
where $u,\,v$ are $N\times N$ matrices and $\Lam$ is $2N\times 2N$.

The scheme of the proof is to express both matrices $\sigma(\vec{X},\psi)$
and $C(\vec{X},\psi)$ in terms of matrices $\sigma(\vec{B}^\pr,\psi)$ and
$C(\vec{B^\pr},\psi)$ and to compare their determinants. Here $\vec{B} =
(A_1,A_2,\ldots,A_N,A^\dg_1, A^\dg_2,\ldots,A^\dg_N) \equiv
(\vec{A},\vec{A^\dg})$ and $\vec{B}^\pr = (\vec{A}^\pr,\vec{A^{\pr\dg}})$.
First we relate $\vec{X}$ to $\vec{B}$,
\be\label{X-B}   
\vec{X}=b\vec{B},\quad
b=\frac{1}{2}\l(\bt{ll}$1_N$&$1_N$\\$i1_N$&$-i1_N$\et\r),
\ee
where $1_N$ is $N\times N$ unit matrix. We introduce $2N\times
2N$ transformation matrix $V$, which relates $\vec{B}$ and $\vec{B}^\pr$,
\be\lb{B-Bpr}   
\vec{B}^\pr = V\vec{B},\quad V=\l(\bt{ll}$u$&$v$\\$v^*$&$u^*$\et\r),
\ee
where $u$ and $v$ are $N\times N$ matrices of the transformation
(\ref{A-Apr}). We consider the new operators $A^\pr_j$ independent (as
well as the old ones $A_j$), therefor matrix $V$ is supposed to be
invertable, that is $\det V \neq 0$. Using the above two linear
transformation and the definition of $\sigma$ we get
\be\lb{sigX-sigA}   
\sigma(\vec{X},\psi)=bV^{-1}\sigma(\vec{A^\pr},\psi)(V^{-1})^Tb^T
\ee
and similarly
\be\lb{CX-CBpr}   
C(\vec{X},\psi)= bV^{-1}C(\vec{B}^\pr,\psi)(V^{-1})^Tb^T.
\ee
Next, using the eigenvalue eqs. (\ref{eve1}) we can prove the equality
\be\lb{detSBpr=detCBpr}   
\det\sigma(\vec{B^\pr},\psi)=\det C(\vec{B^\pr},\psi),
\ee
which in view of (\ref{sigX-sigA}) and (\ref{CX-CBpr}) (and non degenaracy
of $b$ and $V$, $\det b = (-i/2)^N$) leads to the desired equality in the
RUR (\ref{RUR1}),
\be\lb{=RUR1}    
\det\sigma(\vec{X},\psi)= \det C(\vec{X},\psi).
\ee
The proof of auxiliary equality (\ref{detSBpr=detCBpr}) can be carried
out by direct calculations: one has
\begin{eqnarray}   
\sigma_{jk}(\vec{B^\pr},\psi)&=&0\,=\,C_{jk}(\vec{B^\pr},\psi), \nnb\\
\sigma_{N+j,N+k}(\vec{B^\pr},\psi)&=&0\,=\,C_{N+j,N+k}(\vec{B^\pr},\psi),
\nnb\\
\sigma_{j,N+k}(\vec{B^\pr},\psi)&=&-iC_{j,N+k}(\vec{B^\pr},\psi),\nnb\\
\sigma_{N+j,k}(\vec{B^\pr},\psi)&=& iC_{N+j,k}(\vec{B^\pr},\psi),
\end{eqnarray}
which manifestly ensure (\ref{detSBpr=detCBpr}). Thus the states which
satisfy eq. (\ref{eve1}) minimize the inequality (\ref{RUR1}).

States which minimize RUR (\ref{RUR1}) for observables $(X_1,\,X_2,\ldots,
X_n)$ $\equiv  \vec{X}$ should be called {\it Robertson intelligent
states} for $\vec{X}$ (briefly $\vec{X}$-RIS). Equivalent terms could be
Robertson minimum uncertainty states or Robertson correlated states,
following for example papers \ci{183,DKM,Tr93}. However we reserve the
term correlated for states with nonvanishing correlations (covariances)
only.  In case of even $n$ in view of (\ref{eve1}) and (\ref{A-Apr}) RIS
should be denoted as $|\vec{z},u,v\ra$ or $|\vec{z},\Lam\ra$.  For $n=2$
the relation (\ref{RUR1}) coincides with the Schr\"odinger one, eq.
(\ref{SUR}), and RIS are in fact SIS. For two observables the condition
(\ref{eve1}) is necessary and sufficient \ci{Tr94} to get equality in SUR.

Following the analogy to the known case of canonical observables $p_j$ and
$q_j$ one can introduce squeeze operator \ci{Feng,LoKn,XMa} for arbitrary
observables (generalized squeeze operator) $S(u,v)$ as operator which is a
map
from noncorrelated RIS with equal uncertainties for all pairs $X_j$ and
$Y_j=X_{N+j}$
(those RIS minimize Heisenberg relation for $2N$ operators (\ref{HUR})) to
correlated RIS (RIS with nonvanishing covariances and nonequal variances).
Noncorrelated RIS with equal uncertainties for $X_j$ and $Y_j$ are
obtained when $u_{jk}=\delta_{jk}$ and $v_{jk}=0$ in $|\vec{z},u,v\ra$.
\be\lb{S(u,v)}  
S(u,v): \quad |\vec{z},u,v\ra = S(u,v)|\vec{z}\ra,
\ee
where $|\vec{z}\ra = |\vec{z},u\!=\!1,v\!=\!0\ra$. $|\vec{z}\ra$ are
eigenstates of all $A_j$, $j=1,2,\ldots,N$. For two arbitrary observables
the operator $S(u,v)$ was introduced in \ci{Tr94}. This definition is of
importance for generation of RIS $|\vec{z},u,v\ra$ from $|\vec{z}\ra$ when
the states $|\vec{z}\ra$ are known and available.  IS $|z\ra$ with equal
uncertainty  for two observables $X$, $Y$ are constructed, in different
notations, in a number of cases
\ci{GlKl,DKM,BaGi,KuWa,AlGu,SuWaWa}. It is interesting to
note that for certain systems the squeeze operator $S(u,v)$ may exist as
an isometric (not unitary) operator. Such is the case of $S(u,v)$ for the
quadratures of squared boson annihilation operator $a^2$, considered in
\ci{Tr96}. If $S$ is isometric only then its generator $H$ (defined by $S
= \exp(iH)$) is a symmetric (not hermitean  = selfadjoint) operator and
can be considered as generalized observable \ci{TCT}. In such cases
representing $S = \exp(itH)$ ($t$ being real parameter, the time) we see
from (\ref{S(u,v)}) that RIS (for $n=2$ in fact SIS) $| z,u,v\ra$ can be
generated from states with equal uncertainties $| z\ra$ in a process of
nonunitary evolution governed by symmetric Hamiltonian $H$.  Symmetric but
not selfadjoint is e.g. the particle momentum on a half line  and the
Hamiltonian of a particle with different mass parameters in $X$, $Y$ and
$Z$ directions (moving in a crystal) \ci{TCT}.

Now a natural question of existence of RIS arises. We have a positive
answer to this question for a broad class of observables $X_\mu$: RIS
exist for the operators of hermitean representations of semisimple Lie
algebras in Hilbert space $\cal{H}$ and for representations of solvable
algebras $L$ in finite dimensional $\cal{H}$. RIS may exist for infinite
dimensional representations of certain solvable algebras.  The existence
of RIS for any finite dimensional representation of a solvable Lie algebra
$L$ stems from the theorem \ci{Na} that any such representation possess at
least one weight (i.e. a vector exist, which is eigenvector of all
elements of $L$).

\section{Examples of RIS}
\subsection{RIS for Semisimple Lie algebras}

First we note that for any Lie group $G$ the group related CS
\ci{KlSk,Feng} $|\psi(g)\ra = U(g)|\psi_0\ra$ with $|\psi_0\ra$ being
eigenvector of at least one generator $X_\mu$ (these are CS with symmetry)
universally are RIS for the generators of $G$.  Indeed,
$U(g)|\psi_0\ra$ is evidently eigenstate of hermitean operator $U(g)X_\mu
U^\dg(g)$ ($U(g)$ is unitary representation of $G$).  Then we can apply the
proposition 1 and get $\det\sig(\vec{X};\psi(g)) = 0$. Here $\det C$ also
vanishes identically with respect to $g\in G$, i.e.
$\det\sig(\vec{X};\psi(g)) = \det C(\vec{X};\psi(g)) = 0$.
If G is semisimple then hermitean generators $H_l$ from Cartan subalgebra
always have normalizable eigenvectors $|\psi_0\ra$ \ci{BaRo}.  Therefor CS
$|\psi(g)\ra$ with these $|\psi_0\ra$ as reference vector are RIS for all
group generators ( with the trivial minimization: $\det\sig = \det C =0 $
identically with respect to $g\in G$).

We shall prove now that CS $|\psi(g)\ra$ with maximal symmetry are RIS for
the quadrature components of Weyl lowering operators $E_{-k}$ with the
property $\det\sig \geq 0$. The proof consists in application of
proposition 3. The number of quadrature components $X_k,\,Y_k$ of all
$E_{-k}$ is even, denoted by $2n_w$, where $n_w$ is the number of Weyl
operators $E_{-k}$: $E_{-k} = X_k-iX_{n_w+k}\equiv X_k-iY_k$,
$k=1,2,\ldots,n_w$.  We shall prove that the eq.  (\ref{eve1}) ( the
sufficient condition for RIS) is satisfied by CS $|\psi(g)\ra$. As
operators $A_j$ we take here $E_{-k}$ and as $A_j^\pr$ we have to take
linear combinations of Weyl lowering and raising operators
$u_{jk}E_{-k}+v_{jk}E_k$, $j,k = 1,2,\ldots, n_w$ and then consider the
eigenvalue equation
\be\lb{eve2}  
(u_{jk}E_{-k}+v_{jk}E_k)|\vec{z},u,v\ra = z_j|\vec{z},u,v\ra.
\ee
Consider the action of $u_{jk}E_{-k}+v_{jk}E_k$ on the state
$|\psi(g)\ra$.  One has (summation over repeated indices,
$E_k=E_{-k}^\dg$, $H_{l} = H_l^\dg$).
\bear\lb{eve3}   
\l(u_{jk}E_{-k}+v_{jk}E_k\r)|\psi(g)\ra
 = \l(u_{jk}E_{-k}+v_{jk}E_k\r)U(g)|\psi_0\ra \nnb \\
 =U(g)U^{-1}(g)\l(u_{jk}E_{-k}+v_{jk}E_k\r)U(g)|\psi_0\ra \nnb \\
 =U(g)[\l(u_{jk}\tl{u}_{ki}+v_{jk}\tl{v}^*_{ki}\r)E_{-i} +
\l(u_{jk}\tl{v}_{ki}+v_{jk}\tl{u}^*_{ki}\r)E_i \nnb \\
+\l(u_{jk}\tl{w}_{kl}+v_{jk}\tl{w}_{kl}\r)H_{l}]|\psi_0\ra.
\eear
In the above we have applied the BCH formula to the transformations
$U^{-1}E_{k}U$ ($k,j,i = 1,2,\ldots,n_w, \quad l,m=1,2,\ldots,n_c$, $n_c$
being the dimension of Cartan subalgebra)
\be  
U^{-1}(g)E_{-k}U(g) = \tl{u}_{ki}E_{-i} + \tl{v}_{ki}E_i
+\tl{w}_{kl}H_{l}.
\ee
Taking into account that $E_{-i}|\psi_0\ra=0$ and $H_{l}|\psi_0\ra =
h_{l}|\psi_0\ra$ we see that $|\psi(g)\ra$ should be an eigenstate of
all $A^\pr_{j}$ if the $n_w\times n_w$ matrices $u,\,v,\,\tl{u}$ and
$\tl{v}$ satisfy the equation
\be\lb{gcscond}    
u\tl{v} + v\tl{u}^* = 0.
\ee
In the last equation $\tl{u}=\tl{u}(g)$ and $\tl{v}=\tl{v}(g)$ should be
treated as known for a given Lie group representation $U(g)$. Moreover
the matrix $\tl{u}$ is non degenerate. Therefor we always can solve the
eq. (\ref{gcscond}), $v=-u\tl{v}(g)\tl{u}^*{}^{-1}(g)$ and get
$|\psi(g)\ra$ as eigenstate of $A^\pr_{j}=u_{jk}E_{-k}+v_{jk}E_k$,
\be\lb{eve4}   
(u_{jk}E_{-k}+v_{jk}E_k)|\psi(g)\ra = z_j|\psi(g)\ra,
\ee
with eigenvalues $z_j = (u_{jk}\tl{w}_{kl}+v_{jk}\tl{w}_{kl})h_{l}$. In
view of (\ref{eve4}) the group related CS with maximal symmetry
$|\psi(g)\ra$ can be parametrized as RIS for $2n_w$ components of Weyl
generators: $|\psi(g)\ra = |\vec{z},u,v\ra$ where $u$, $v$ are $n_w\times
n_w$ matrices.

Thus we have demonstrated that states from unitary (in particular unitary
and irreducible) orbits of exstremal weight vectors of semisimple Lie
algebras are RIS for all basis operators $X_\mu$ and for the quadratures
$X_k,\,Y_k=X_{n_w+k}$ of Weyl operators $E_{-k}$ as well.  As far as we
know this intelligence property of the group related CS wasn't noted so
far in the literature.

We underline that RIS for quadrature components of Weyl generators
$E_{-k}$ are more general than the group related CS with maximal symmetry:
states $|\psi(g)\ra$ are only a part of the set of solutions of eigenvalue
eq.  (\ref{eve2}), corresponding to the constrain (\ref{gcscond}) on the
parameters $u_{jk}$ and $v_{jk}$.  On the example of $su(1,1)$ and $su(2)$
($n_w=1,\,\, n_c=1$) this was analyzed by explicit constructions of SIS
$|z,u,v;k\ra$ in ref. \ci{Tr94}.

It is worth noting that the propositions 1 and 3  can be applied to any
subset of the operators of a given Lie algebra $L$. Therefor it makes
sense to consider the eigenvalue problem for general element of the
complexified algebra $L^C$,
\be\lb{acs}  
(\beta_{\nu}X_\nu)|\psi\ra = z|\psi\ra,
\ee
where $X_\nu$ ($\nu=1,\ldots,n$) are basis operators of $L$ and $\beta_\nu$
are complex parameters.  Taking specific constrains on the complex
parameters $\beta_\nu$ one can get various subset of RIS for less than $n$
algebra operators, in particular various $X_j$-$Y_k$ SIS. The
property of group related CS to be part of the set of eigenstates of
complex linear combinations of all algebra operators was noted in
\ci{Br96,Tr96}. States that satisfy (\ref{acs}) could be called algebraic
CS \ci{Tr96} or algebra eigenstates \ci{Br96}.

\subsection{Explicit solutions for $su(1,1)$ and $su(2)$ RIS}

Consider first $su(1,1)$ case. The basis elements of $su(1,1)$ are three
operators $K_\mu$, $\mu=1,2,3$, which obey the relations
\be\lb{su11cr}  
[K_1,K_2]=-iK_3,\quad [K_2,K_3]=iK_1, \quad [K_3,K_1]=iK_2.
\ee
The Casimir operator is $C_2 = K_3^2-K_2^2-K_1^2 = k(k-1)$ and Weyl
lowering and raising operators are $K_{\mp}= K_1\mp iK_2$.  According to
the previous discussion RIS for all three algebra operators and for any
pair $K_j$-$K_k$ are contained in the set of eigenstates of general
element of the algebra.  Therefor one has to consider the eigenvalue
equation for the general element of $su^C(1,1)$,
\be\lb{su11acs}  
(uK_- + vK_+ + wK_3)| z,u,v,w;k\ra = z| z,u,v,w;k\ra,
\ee
where $u,v,w$ are complex parameters, simply related to
$\beta_\nu$ introduced in  (\ref{acs}).
 This equation can be solved \ci{Tr96,Br96} using the Barut-Girardello CS
representation (BG representation) \ci{BaGi} or the $SU(1,1)$ group
related CS representation \ci{KlSk,Feng}). The solution can be carried out
for $su(1,1)$ representations with Bargman index $k = 1/4,\,3/4$ and for
the discrete series $k = 1/2,\,1,\,3/2,\ldots$  (patricular cases of
$v=0=w$ and $w=0$ were solved in \ci{BaGi,Tr94}. The Barut-Girardello CS
(BG CS) $|\eta;k\ra$ are eigenstates of $K_-$: $K_-|\eta;k\ra =
\eta|\eta;k\ra$. In this representation
\be\lb{BGrep}  
K_+ = \eta,\,\quad
K_- = 2k\frac{d}{d\eta}+\eta\frac{d^2}{d\eta^2},\,\quad
K_3 = k+\eta\frac{d}{d\eta},
\ee
and states $|\psi\ra$ are represented by analytic functions $\Phi(\eta)$
which up to a certain common factor $f(|\eta|)$ are proportional to $\la
k;\eta^*|\psi\ra$. Orthonormalized eigenstates $| m;k\ra$ of $K_3$ are
represented by monomials $\eta^m\,[\Gamma(k)/(m!\Gamma(m+k))]^{1/2}$. For
$u\neq 0$ the required analytic solution of (\ref{su11acs}) is
\ci{Tr96}
\be\lb{Phi_z}   
\Phi_z(\eta;u,v,w) = N(z,u,v,w)\exp(c\eta)M(a,b,c_1\eta)
\ee
where $N(z,u,v,w)$ is a normalization constant, $M(a,b,\eta)$ is the
Kummer function (confluent hypergeometric function $_1F_1(a,b;\eta)$)
\ci{Stegun}, parameters $a,\,b,\,c$ and $c_1$ are
\bear\lb{abc}  
a = k+\frac{z}{\sqrt{w^2-4uv}},\,\,\quad b=2k,\nonumber\\
c = -\frac{1}{2u}\l(w+\sqrt{w^2-4uv}\r),\quad c_1=\frac{1}{u}
\sqrt{w^2-4uv},
\eear
and the normalizability conditions take the form
\be\lb{ncond}   
\frac{1}{2| u|}\l| w-\sqrt{w^2-4uv}\r| \,\, < \, 1,
\quad {\rm or} \quad
\frac{1}{2| u|}\l| w+\sqrt{w^2-4uv}\r| \,\, < \,1.
\ee
When the inequalities (\ref{ncond}) are broken down the functions
$\Phi_z(\eta;u,v,w)$  still are solutions of eq. (\ref{su11acs}) and could
be considered as non normalizable eigenstates. In case of $u=0$ in eq.
(\ref{su11acs}) we have (in view of (\ref{BGrep}) first order equation to
solve \ci{Tr96}. It turned out that the solutions for this case could be
obtained from $\Phi_z(\eta;u,v,w;k)$ taking appropriate limits in it.  One
can check that the conditions (\ref{ncond}) can be satisfied by
real $w$ and $v=u^*$ when the operator $uK_- + vK_+ + wK_3$ becomes
hermitean. Then the algebraic states $| z,u,u^*,w\ra$ ($w=w^*$) are RIS for
the three observables $K_1,\,K_2$ and $K_3$. RIS for the nonsquare
integrable representations corresponding to $k=1/2,\,3/4$ are considered
in next subsection.

Various known states are contained in the large family of $su(1,1)$ states
$| z,u,v,w;k\ra$ \ci{Tr96}. In particular, when $w=0$ we get the
SIS $| z,u,v;k\ra$ for the noncompact generators $K_1$ and
$K_2$, which in turn at $ z=-k\sqrt{-uv}$ \ci{Tr94} recover the family of
$SU(1,1)$ group CS $|\tau;k\ra$ (the squeezed vacuum states \ci{Feng}),
$\tau=\sqrt{-v/u},\,|\tau|<1$. In view of
the positivity of the commutator $i[K_1,K_2] = [K_-,K_+]/2$ the
uncertainty matrix $\sig(K_1,K_2;\rho)$ is positive definite and therefor
possesses the resulting properties, described in section 2. In IS
$| z,u,v;k\ra$ the matrix elements of $\sig$ are
\be\lb{sig2}  
\sig_{11} = \frac{1}{2}\frac{| u-v|^2}{| u|^2-|v|^2}\la K_3\ra,\quad
\sig_{22} = \frac{1}{2}\frac{| u+v|^2}{| u|^2-|v|^2}\la K_3\ra,\quad
\sig_{12} =  \frac{{\rm Im}(u^*v)}{| u|^2-| v|^2}\la K_3\ra,
\ee
satisfying $\det\sig = \det C = \la K_3\ra^2/4$. The $K_1$-$K_3$ and
$K_2$-$K_3$ IS are obtained from $| z,u,v,w;k\ra$ when $v=u$ and $v=-u$
respectively.

The case of $su(2)$ RIS (i.e. spin RIS) can be treated in a similar manner
using the representation of $SU(2)$ group related CS $|\zeta;j\ra$  in
which \ci{Feng}
\be\lb{su2op}  
J_- = -\zeta^2\frac{d}{d\zeta}+2j\zeta,\quad J_-=\frac{d}{d\zeta},\quad
J_3 = \zeta\frac{d}{d\zeta} - j.
\ee
Here $j=1/2,\,1,\,3/2,\ldots$, $\quad [J_-,J_+] = -2J_3$, $[J_3,J_{\pm}] =
\pm J_{\pm}\quad$ ($J_\pm= J_1\pm iJ_2$) and $J^2=J_1^2+J_2^2+J_3^2 =
j(j+1)$. The required eigenvalue problem
\be\lb{su2acs}   
(\beta_\nu J_\nu)| z,\vec{\beta};j\ra = z| z,\vec{\beta};j\ra
\ee
where $\vec{\beta} = (\beta_1,\beta_2,\beta_3)$, ( $\beta_\nu$ are complex
parameters) is solved by Brif \ci{Br96}. In case of $b^2 \equiv
\vec{\beta}\vec{\beta}\neq 0 \neq \beta_1-i\beta_2\equiv \beta_-$ the
solution is \ci{Br96}
\be\lb{su2Phi_z}  
\Phi_z(\zeta;\vec{\beta},j) =
N_0\l(\zeta-\frac{\beta_3-b}{\beta_-}\r)^{j+z/b}
\l(\zeta-\frac{\beta_3+b}{\beta_-}\r)^{j-z/b},
\ee
with the normalizability condition $z = mb$, $m=-j,-j+1,\ldots,j-1,j$. As
we expect these $su(2)$ RIS contain the set of standard $SU(2)$ CS with
maximal symmetry $|\zeta^\pr;j\ra$ and this occurs when $m=\pm j$
with $\zeta^\pr = -\beta_-(\zeta_3\mp b)^{-1}$ \ci{Br96}. At $\beta_3=0$
the $su(2)$ RIS coincide with the Schr\"odinger $J_1$-$J_2$ IS considered
in ref. \ci{Tr94}.

For the $su(2)$ observables $J_\nu$ (the spin components) it is important
to note that the spin component uncertainty matrix $\sig(\vec{J};\rho)$ in
any state can be diagonalized by means of orthogonal linear transformation
of $J_\nu$. The latter can be induced by rotation of coordinates in
I$\!\!$R$_3$ since $su(2) \sim so(3)$. Therefor we deduce that spin
component correlations are of pure coordinate nature -- they can be
eliminated in any state by rotations of the reference frame.  Here one can
also perform second kind diagonalization of $\sig$, keeping $J_\nu$ and
transforming the state $\rho$ by an unitary operator $U(g)$ of $SU(2)\sim
SO(3)$. Thus correlated spin RIS are unitary equivalent to noncorrelated
spin RIS.

\subsection{RIS of multimode boson systems}

In this subsection we first consider $n=2N$  canonical operators $p_j$ and
$q_j$, $j = 1,2,\ldots,N$, which are quadrature components of $N$
boson/photon destruction (creation) operators $a_j = (q_j+ip_j)/\sqrt{2}$
($a^\dg_j = (q_j-ip_j)/\sqrt{2}$): $[q_j,p_k] = i\delta_{jk}$. Here we put
for concreteness $X_\nu \equiv Q_\nu$, $Q_j = p_j$, $Q_{N+j} = q_j$ and
$A_j = q_j+ip_j = a_j\sqrt{2}$. The set of $Q_\mu$ and the unity operator
close the Heisenberg algebra $h_N$, which is nilpotent (therefor non
semisimple). So RIS for canonical observables $Q_\mu$ are $h_N$ RIS (to be
called also multimode amplitude RIS).  According to the proposition 3
eigenstates $|\vec{\alf},u,v\ra\equiv|\vec{\alf},\Lam\ra$,
$\quad\vec{\alf} = (\alf_1,\alf_2,\ldots,\alf_N)$,  of $a_j^\pr$,
\be\lb{apr}  
a_j^\pr = u_{jk}a_k+v_{jk}a^\dg_{jk}
=\frac{1}{2}[(\lam_q)_{jk}q_k + (\lam_p)_{jk}p_k],
\ee
with any $u$ and $v$  are RIS for $Q_\mu$,
\be\lb{eve5}   
a^\pr_j|\vec{\alf};u,v\ra = \alf_j|\vec{\alf};u,v\ra,\quad
j=1,2,\ldots,N.
\ee
Here $u=(\lam_q-i\lam_p)/2$, $v=(\lam_q+i\lam_p)/2$ and $u,\,v,\,\lam_q$
and $\lam_p$ are $N\times N$ complex matrices.  The $N\times N$ matrices
$\lam_q$ and $\lam_p$ are related to the transformation matrix $\Lam$ in
(\ref{LT1}) (which now is rewritten as $Q^\pr_\mu = \lam_{\mu\nu} Q_\nu$)
as follows:
\be\lb{Lam}   
\Lam =\l(\bt{ll}
$\lam_1$&$\lam_2$\\$\lam_3$&$\lam_4$\et\r),\quad
\lam_p=\lam_3+i\lam_1,\,\lam_q=i\lam_2+\lam_4.
\ee
If one impose the symplectic conditions $\Lam J\Lam^T = J$ on $\Lam$ the
operators $a^\pr_j$ become new annihilation operators, i.e. the linear
transformation  (\ref{LT1}) becomes canonical one. With this conditions
$\vec{Q}$-RIS are unitary equivalent (with methaplectic operator
$U(\Lam)$) to eigenstates of $a_j$, i.e. to canonical multimode CS
$|\vec{\alpha}\ra$. In ref.
\ci{MMTHo} states $|\vec{\alpha},\Lam(t)\ra$ were constructed explicitly
as solution $|\vec{\alf};t\ra$ of time dependent Schr\"odinger equation
for general quadratic, possibly time dependent, Hamiltonian
$H=B_{\mu\nu}(t)Q_\mu Q_\nu$ (plus linear terms as well). In terms of
parameter matrices $\lam_q$ and $\lam_p$ these canonical RIS
$|\vec{\alf},\Lam\ra = |\vec{\alf},u,v\ra$ in coordinate representation
read ($\Delta = 0$ in eqs. (17) of third paper of ref. \ci{MMTHo})
\be\lb{Q-RIS}    
\la \vec{q}|\vec{\alf};\Lam\ra = \pi^{N/4}\exp\l(\gamma +
\vec{\tl{\nu}}\vec{q}
-\frac{1}{2} \vec{q}\tl{\mu}\vec{q}\r),
\ee
where $\tl{\mu}$ is $N\times N$ matrix, $\tl{\mu}=i\lam^{-1}_p\lam_q$,
$\vec{\tl{\nu}}$ is $N$ vector,
$\vec{\tl{\nu}}=(1/\sqrt2)\l(\lam_q^\dg-\lam_p^{-1}\lam_q\lam_p^\dg\r)
\vec{\alf}$  and
 $$\gamma = -\frac{1}{2}|\vec{\alf}|^2 +
\frac{i}{4}\vec{\alf}(\lam_p^*\lam_p^{-1}\lam_q\lam_p^\dg -
\lam_q^*\lam_p^\dg) \vec{\alf}.$$
At $\Lam =1$ the RIS (\ref{Q-RIS}) coincide with canonical CS
$|\vec{\alf}\ra$ in coordinate representation. The multimode states
(\ref{Q-RIS}) in deferent parametrizations were also considered in
several papers under the names multimode squeezed states \ci{XMa} or
multimode/polymode correlated states \ci{SCB-Tr95,DoMaMa,Sud2} or Gaussian
pure states \ci{Sud2}.

It is worth noting that for canonical RIS the condition
(\ref{eve5})$\,\sim\,$(\ref{eve1}) is not only sufficient, but also
necessary, i.e. all $\vec{Q}$-RIS are eigenstates of $a_j^\pr = u_{jk}a_k
+ v_{jk}a^\dg_{jk}$ for some $u_{jk}$ and $v_{jk}$. This can be proved
using the diagonalization of $\sig(\vec{Q},\rho)$.  Indeed, let $\sig^\pr
= \sig(\vec{Q^\pr},\rho) = \sig(\vec{Q},\rho^\pr)$ be diagonal. Here
$\rho^\pr = U(\Lam)\rho U^\dg(\Lam)$, where $U$ is methaplectic unitary
operator. Since the mean commutator matrix $C$ now is constant, $\det C =
4^{-N}$ the equality in RUR is $\det\sigma = \det\sigma^\pr=\prod_j^N
\sig^\pr_{jj}\sig^\pr_{N+j,N+j}=4^{-N}.$
Since for every $j$ the product $\sig^\pr_{jj}\sig^\pr_{N+j,N+j}$ is
greater or equal to $1/4$ we get that all products should be equal to
$1/4$.  But this is possible if and only if $\rho$ is pure multimode CS
for new variables $\vec{Q^\pr}$, that is $\rho$ is pure state,
methaplectically equivalent to multimode CS for old variables $\vec{Q}$,
$\rho = U(\Lam)|\vec{\alf}\ra\la\vec{\alf}| U^\dg(\Lam)$.

Consider briefly now the uncertainty matrix of canonical observables
$\sigma(\vec{Q},\rho)$. Since $Q_\mu$ satisfy the requirements of
proposition 2 the $\sigma(\vec{Q},\rho)$ is positive definite. Therefor it
can be diagonalized by means of linear canonical transformation in any
state $\rho$ and it obey the inequalities (\ref{newur3}).  In
$\vec{Q}$-RIS $|\vec{\alf},\Lam\ra$ the dispersion matrix
$\sigma(\vec{Q};\vec{\alf},\Lam)$ has further properties. The main one is
that $\sigma(\vec{Q};\vec{\alf},\Lam)$ is symplectic itself. Indeed we
have
\be\lb{sig3}  
\sigma(\vec{Q};\vec{\alf},\Lam)= \sigma(\vec{Q^\pr};\vec{\alf},1) = \Lam
\sigma(\vec{Q};\vec{\alf},1)\Lam^T,
\ee
where $\sigma(\vec{Q};\vec{\alf},1)$ is the uncertainty matrix in
multimode canonical CS $|\vec{\alf}\ra$. The latter is evidently
proportional to the unity, $\sigma(\vec{Q};\vec{\alf},1)= \frac{1}{2}$
and therefor if $\Lam$ is symplectic then $2\sig(\vec{Q};\vec{\alf},\Lam)$
is also symplectic. We express $\sigma$ in terms of $N\times N$
uncertainty matrices $\sigma_{pp}$, $\sigma_{qq}$, $\sig_{qp}$ and
$\sig_{pq} = \sig_{qp}^T$
\be\lb{sig4}  
\sigma(\vec{Q})=\l(\bt{ll}$\sig_{pp}$&$\sig_{pq}$\\
$\sig_{qp}$&$\sig_{qq}$\et\r)
\ee
and write the symplectic properties of $\sig(\vec{Q};\vec{\alf},\Lam)$ in
$N\times N$ matrix form,
\be\lb{sig5}  
\sig_{pp}\sig_{qp}-\sig_{pg}\sig_{pp}=0=\sig_{qp}\sig_{qq}-\sig_{qq}
\sig_{pq}, \quad \sig_{pp}\sig_{qq}-\sig_{pq}^2=1/4.
\ee
For $N=1$ the last equality is just the equality in Schr\"odinger relation
(\ref{SUR}), the first two being satisfied identically in any state.

For boson systems it is of interest to consider observables which are
quadratic combinations of creation and annihilation operators $a^\dg_j$
and $a_k$ (or equivalently of $p_j$ and $q_k$). Quadratic combinations
\be\lb{spNop}  
K_{jk}=\frac{1}{2}a_ja_k, \quad K_{jk}^\dg=\frac{1}{2}a^\dg_ka^\dg_j,
\quad K^{(3)}_{jk} = \frac{1}{4}(a^\dg_ja_k + a^\dg_ka_j)
\ee
close the simple noncompact algebra $sp(N,R)$ \ci{BaRo}, the noncompact
elements being spanned by lowering and raising operators $K_{jk}$ and
$K_{jk}^\dg$. In the one mode case $sp(1,R)\sim su(1,1)$ and
\be\lb{sp1op}  
\frac{1}{2}a^2 = K_-,\quad \frac{1}{2}a^{\dg 2} = K_+,\quad
\frac{1}{4}(2a^{\dg 2}a +1) = K_3.
\ee
Operators (\ref{spNop}) are generators of the methapletic group $Mp(N,R)$,
which covers the $Sp(N,R)$.  $sp(N,R)$ RIS in the representation
(\ref{spNop}) should be called multimode squared amplitude RIS. RIS for
the quadratures $X_{jk}$ and $Y_{jk}$ of $K_{jk}$, $K_{jk} = X_{jk} +
iY_{jk}$ (shortly $K_{jk}$-RIS), are eigenstates of $N\times N$ complex
combinations of lowering and raising operators $K_{jk}$ and $K_{jk}^\dg$
and according to our general result they contain group related $Mp(N,R)$
CS with maximal symmetry, $|\psi(g)\ra=U(g)|0\ra$, $U(g)\in Mp(N,R)$, the
exstremal vector being the multimode boson vacuum $|\vec{0}\ra$ (these CS
coincide with multimode squeezed vacuum states \ci{SCB-Tr95,XMa,DoMaMa}).
On the other hand $Mp(N,R)$ CS are annihilated by all $a^\pr_j =
U(g)a_jU^{-1}(g) = u_{jk}a_k + v_{jk}a^\dg_k$. Herefrom we get
the property that $Mp(N,R)$ CS with maximal symmetry are simultaneously
$h_N$ and $sp(N,R)$ RIS (i.e. amplitude and squared amplitude multimode
RIS, double IS). In coordinate representation and in parametrization by
$\lam_q$ and $\lam_p$ ($\vec{a}^\pr = \lam_q\vec{q}+\lam_p\vec{p}$) these
multimode double IS are given by formula (\ref{Q-RIS}) with
$\vec{\alf}=0$.

An other explicit example of $sp(N,R)$ RIS is given by multimode squeezed
Fock states $U(g)|\vec{n}\ra$, where $U(g) \in Mp(N,R)$. Indeed, Fock
states $|\vec{n}\ra$ are eigenstates of hermitean $Mp(N,R)$ generators
$K^{(3)}_{jj} = a^\dg_ja_j/2$ (see eq. (\ref{spNop})), therefor
$U(g)|\vec{n}\ra$ are eigenstates of hermitean operators
$U(g)K^{(3)}_{jj}U(g)^{\dg}$ which are real linear combinations of all
$Mp(N,R)$ generators (follows from the BCH formula). From section 3 we
know that this eigenvalue property is sufficient for the equalities
$\det\sig = \det C = 0$, i.e.  the squeezed Fock states are $sp(N,R)$ RIS
for all hermitean quadratures of operators (\ref{spNop}).  Multimode
squeezed Fock states $U(g)|\vec{n}\ra$ were constructed in the last two
papers of ref. \ci{MMTHo}, where the $Mp(N,R)$ operator  $U(g)$ was taken as
evolution operator $U(t)$ of general $N$ dimensional quadratic quantum
system (in coordinate representation the states
$\la\vec{q}|U(g)|\vec{n}\ra$ were expressed as product of
$\la\vec{q}|\vec{0}\ra$ (see eq. (\ref{Q-RIS})) and a Hermite polynomial
of $N$ variables). Note that squeezed Fock states are $sp(N,R)$ RIS and
not $h_N$ RIS and squeezed Glauber CS are $h_N$ RIS and not $sp(N,R)$ RIS.
Only squeezed vacuum states are simultaneously $sp(N,R)$ RIS and $h_N$ RIS
($h_N$ RIS = $\vec{Q}$-RIS).

Recently attention is paid in the physical literature to multimode even
and odd CS \ci{Olg} $|\vec{\alf}\ra_\pm = N_\pm(|\vec{\alf}\ra \pm
|-\vec{\alf}\ra)$, where $|\vec{\alf}\ra = D(\vec{\alf})|\vec{0}\ra$ is
Glauber multimode CS. We readily see that these $|\vec{\alf}\ra_\pm$ are
eigenstates of all $K_{jk}$, eq.  (\ref{spNop}), and therefor are
noncorrelated squared amplitude RIS with equal uncertainties of
quadratures of $K_{jk}$. It is the set of all $sp(N,R)$ $K_{jk}$-RIS which
is a natural extension of that of multimode even and odd CS
$|\vec{\alf}\ra_\pm$, incorporating also the multimode squeezed vacuum
states $|\vec{0},u,v\ra$ and Glauber CS $|\vec{\alf}\ra$.
Unlike the even and odd CS $|\vec{\alf}\ra_{\pm}$
the $K_{jk}$-RIS (being eigenstates of combinations $u_{jk}a_ja_k +
v_{jk}a^\dg_ja^\dg_k$) can exhibit strong squeezing in quadratures of
$a_ja_k$ and therefor can be called {\it multimode squared amplitude
squeezed states} in complete analogy to the well known case of multimode
(amplitude) squeezed states \ci{SCB-Tr95,XMa,DoMaMa}.

We underline that the set of all $sp(N,R)$ RIS, and even the set of the
$K_{jk}$-RIS is much larger than the set of $Mp(N,R)$ CS $U(g)| 0\ra$. The
problem can be solved entirely in the one mode case, $N=1$, using Glauber
CS representation, in which $a=d/d\alf$, $a^\dg = \alf$
\ci{Br96,Tr96}. The resulting even states take the form (\ref{Phi_z}) with
the replacements $k=1/4$ and $\eta = \alf^2/2$, the normalizability
conditions remaining the same as (\ref{ncond}). Some particular sets of
one mode squared amplitude squeezed states are constructed and discussed
in
\ci{BeHiYu}. Generalized one mode even and odd CS $|z,u,v;\pm\ra$ were
first constructed in the second paper of ref.
\ci{SCB-Tr95} as even and odd solutions of the eigenvalue equation
$(ua^2+va^{\dg 2})|z,u,v;\pm\ra = z|z,u,v;\pm\ra$ with complex parameters
$u$ and $v$. Eigenvalue problem for operators $(a+\zeta a^\dg)^2$ ($\zeta
\in C\!\!\!\!l$ was considered in \ci{Wun}.

The RIS which are not group related CS exhibit many physical properties
which group CS lack. One of such properties is squeezing in fluctuation of
group generators. Squeezing in fluctuation of $X_\mu$ in a state
$|\psi\ra$ occurs if $|\psi\ra$ is close (by norm form example) to an
eigenstate of $X_\mu$ since the (squared) variance $\Delta^2
X_\mu=\sig_{\mu\mu}$ of $X_\mu$ vanishes in eigenstates of $X_\mu$ {\it
only} \ci{Tr94}.  Therefor if in RIS which is eigenstate of $\beta_\nu
X_\nu$ all but $\beta_\mu$ tend to $0$ then $\Delta X_\mu$
should tend to $0$. In group CS with symmetry it is not always possible to
let all but $\beta_\mu$ to tend to $0$ due to constrain (\ref{gcscond})
(it is trivially possible if $X_\mu$ itself has $|\psi_0\ra$ as its
eigenstate).  In the case of $su(1,1)$ we have explicit solutions
$| z,u,v,w;k\ra$, eq.  ({\ref{Phi_z}) and CS $|\tau;k\ra$ and one
can verify
the above statement: the variances of $K_{1,2}$ in CS are grater than
$k$ for any $\tau$ \ci{Tr94}, while for example for $k=1/4$ the
$K_1$-$K_2$ IS $| z,u,v;1/4\ra$  with $z=-1,\,u=\sqrt{1+x^2},\,v=-x<0$
exhibit strong squeezing in $K_2$ ($\Delta K_2$ is monotonically
decreasing when $x$ increases). Moreover one can find IS which
exhibit $K_1$ ($K_2$) and $q$ ($p$) squeezing (joint amplitude and squared
amplitude squeezing) simultaneously. Subpoissonian statistics also occurs
in IS $| z,u,v,w;1/4\ra$. In greater detail nonclassical properties of
$SU(1,1)$ IS (for $k=1/4,\,3/4$) are discussed (and illustrated by several
graphics) in \ci{Tr96}.

By means of $4$ boson operators $a,\,b,\,a^\dg,\,b^\dg$ one can construct
quadratic combinations which close $su(1,1)$ (the representations with
Bargman index $k = (1 + |n_a-n_b|)/2 = 1/2,1,\ldots$,
considered in the previous subsection) or $su(2)$ algebra
(the Schwinger realization), which are subalgebras of $sp(4,R)$, eq.
(\ref{spNop}) for $N=2$.  Currently physical properties of various
$su(1,1)$ and $su(2)$ SIS of two mode boson/photon system are being
discussed (see \ci{LuPe,BrAr,BrMa} and references therein). We note the
result of \ci{BrMa}: $K_2$-$K_3$ two mode IS which are not $SU(1,1)$ group
CS can improve the sensitivity in the interferometric measurements.
Several schemes of generation of SIS for $su(1,1)$ or $su(2)$ operators in
two mode quadratic boson representations are considered recently
\ci{LuPe,BrAr,BrMa}. But so far no scheme of generation of $K_1$-$K_2$ one
mode SIS is presented.  It seems natural to generate these SIS from
experimentally available Glauber CS or even and odd CS \ci{DKM} acting on
the laster by the squared amplitude squeeze operator $S(u,v)$, eq.
(\ref{S(u,v)}). For this purpose however one has to look for a nonunitary
evolution process, since here the squeeze operator $S(u,v)$ is isometric
only \ci{Tr96}.

\section{Concluding remarks}

We have shown that the uncertainty matrix for $n$ observables $X_\mu$ can
always be diagonalized by linear transformation of $X_\mu$. For the case
of spin component operators this means that spin covariances are of pure
coordinate origin and correlated spin states are unitary equivalent to
noncorrelated states. When the uncertainty matrix is positive definite (as
is the case e.g. of $q$-deformed multimode boson system with $q > 0$, in
particular the case  of canonical boson system, $q = 1$) it can be
diagonalized by means of symplectic transformations. Using the above
diagonalization property a new family of uncertanty relations for positive
definite uncertainty matrices is established.

The Robertson $n$ dimensional relation for the uncertainty matrix, eq.
(\ref{RUR1}), is shown to be efficient in generalization of the basic
properties of Glauber coherent states (CS) to arbitrary system of
observables $X_\mu$. For even number $n$ of observables this relation is
minimized in a state $|\psi\ra$ if $|\psi\ra$ is eigenstate of $n/2$
independent complex combinations of $X_\mu$. For any (even or odd) $n$ the
minimization occurs in states which are eigenvectors of real combination of
$X_\mu$. When $X_\mu$ close a semisimple Lie algebra the set of states
which minimize the Robertson inequality (called here Robertson intelligent
states (RIS)) contain the corresponding group related CS with symmetry. CS
with maximal symmetry are contained also in RIS for the quadratures of
Weyl lowering and raising operators. Thus it is the Robertson uncertainty
relation  that brings together the three ways of generalization of Glauber
CS \ci{Feng} to the level of $n$ observables.

RIS which are not group related CS can exhibit interesting physical
properties. One such universal property to be distinguished from CS is the
strong squeezing of group generators. In this way the multimode squared
amplitude squeezed states are naturally introduced as $sp(N,R)$ RIS.
Squared amplitude  RIS can exhibit both linear and quadratic squeezing as
we have shown on the example of $K_1$-$K_2$ IS.  Such joint squeezing of
noncommuting observables could be useful in optical communications and
interferometric measurements since the field in such squeezed states is
better determined - this should be considered elsewhere.  The problem of
generation of RIS for two $su(1,1)$ and $su(2)$ observables is discussed
in recent papers \ci{LuPe,BrAr,BrMa}.  In this connection we note the
principle possibility to generate e.g. $K_1$-$K_2$ squared amplitude IS
by means of isometric (non unitary) evolution operators.

{\bf Acknowledgement}. The work is partially supported by Bulgarian
Science Foundation, Contracts No. F-559.


\begin{thebibliography}{99}
\baselineskip 18pt

\bibitem{GlKl}  Glauber R~J~ 1963, Phys. Rev. {\bf 131} 1726 \\[-1mm]
		  Klauder J~R 1963, J. Math. Phys. {\bf 4} 1005.
\bibitem{KlSk}   Klauder J~R and   Skagerstam B~S 1985. {\it
		Coherent States} (W. Scientific, Singa-\\[-1mm]pore, 1985).
\bibitem{Feng}   Zhang W~M,  Feng D~H  and Gilmore R 1990, Rev. Mod.
		Phys. {\bf 62}, 867.
\bibitem{Ro}  Robertson H~R 1934,  Phys. Rev. {\bf 46} 794.
\bibitem{183}  Dodonov V~V and  Man'ko V~I 1987, in Trudy FIAN, v. 183,
        p.1-70 ("Nauka",\\[-1mm] Moskva, 1987) (Nuova Science,  Commack,
		N.Y., 1988).
\bibitem{BaRo}   Barut A~O and Raszcka R 1977. {\it Theory of Group
        Representations and \\[-1mm]Applications} (Polish
		Publishers, Warszawa, 1977).
\bibitem{Ar} Aragone C, Chalband E and Salamo S 1976, J. Math. Phys. {\bf
        17} 1963.
\bibitem{DKM} Dodonov V~V, Kurmyshev E and Man'ko V~I 1980, Phys. Lett.
		A{\bf 79} 150.
\bibitem{MMT69} Malkin I~A, Man'ko V~I and  Trifonov D~A 1969,  Phys.
        Lett. A{\bf 30} 413\\[-1mm]
\bibitem{MMTStoLu} Malkin I~A, Man'ko V~I and  Trifonov D~A 1970 		
		 Phys. Rev. D{\bf 2} 1371\\[-1mm]
	 	 Stoler D~A 1970, Phys. Rev. D{\bf	1} 3217\\[-1mm]
		 Lu E~Y~C 1971, Lett. N. Cimento {\bf 2}, 1241.
\bibitem{MMTHo} Holz A 1970, Lett. N. Cimento A{\bf 4} 1319\\[-1mm]
		Malkin I~A, Man'ko V~I and  Trifonov D~A 1971,
									N. Cimento  A{\bf 4} 773\\[-1mm]
		-- 1973, J. Math. Phys. {\bf 14} 576.
\bibitem{Sch} Schr\"odinger E 1930, in Sitz. der Preuss.  Acad. Wiss.
        (Phys.-Math. Klasse,\\[-1mm] p.296) (Berlin, 1930).
\bibitem{Tr93} Trifonov D~A 1993, J. Math. Phys. {\bf 34} 100.
\bibitem{LoKn} Loudon R and Knight P 1987, J. Mod. Opt. {\bf 34}, 709\\[-1mm]
        Walls D~F 1983, Nature {\bf 306} 141.
\bibitem{Tr94} Trifonov D~A 1994a, J. Math. Phys. {\bf 35},  2297\\[-1mm]
        -- 1994b, Phys. Lett. A{\bf 187} 284.
\bibitem{BaGi}  Barut A~O and Girardello L 1971, Commun. Math. Phys.
        {\bf 21} 41.
\bibitem{BeHiYu} Hillery M 1984,  Phys. Rev.  A{\bf 36} 3796\\[-1mm]
		 Bergou J~A, Hillery M  and  Yu D 1991,  Phys.  Rev.  A{\bf
		43}, 515\\[-1mm]
		Nieto M~M and Truax D~R, Phys. Rev. 1993, Lett. {\bf 71} 2843\\[-1mm]
        Gerry C~C and Grobe R 1995, Phys. Rev. {\bf 51} 4123\\[-1mm]
		Puri R~R and Agarwal G~S 1996, Phys. Rev.A{\bf 53}(3), 1786.
\bibitem{ProVal} Provost J and Vall\'ee G 1980, Commum. Math.
		Phys. {\bf 76} 289\\[-1mm]
		Nikolov B~A and Trifonov D~A 1988, Bulg. J.	Phys. {\bf 15}, 33
		\\[-1mm] Abe S 1993, Phys. Rev. A{\bf 48}, 4102\\[-1mm]
		Spera M 1993, J. Geom. Phys. {\bf 12}, 165.
\bibitem{SCB-Tr95}  Sudarshan E~C~G, Chiu C~B and Bhamathi G
		1995, Phys. Rev. A{\bf 52}, 43\\[-1mm]
		Trifonov D~A 1995, Preprint INRNE-TH-95/5.
\bibitem{Ga} Gantmaher F~R 1975. {\it Teoria Matrits} (Moskva, "Nauka",
		1975).
\bibitem{CoBoGo} Colpa J~H 1978, Physica {\bf 93}A, 327\\[-1mm]
     	Bogdanovic R and Gopinathan M 1979, J. Phys. A{\bf 12} 1457.
\bibitem{Se} Seleznyova A~N 1995, Phys. Rev. A{\bf 51} 950.
\bibitem{MaBi} Macfarlane A~J 1989,  {\it J. Phys.} A{\bf 22} 4581\\[-1mm]
		Biedenharn L~C 1989, {\it J. Phys.} A{\bf 22} L873.
\bibitem{Oh} Oh C~H and Singh K 1994, J. Phys. A{\bf 27} 5907.
\bibitem{KuWa} Kuang L~M and Wang F~S 1993, Phys. Lett. A{\bf 173} 221.
\bibitem{AlGu} Aldaya V and Guerrero J 1995, J. Math. Phys. {\bf 36} 3191.
\bibitem{SuWaWa} Agarwal G~S 1988, J. Opt. Soc. Am. B{\bf 5} 1940\\[-1mm]
		Sun J, Wang J and Wang C  1991, Phys. Rev. A{\bf 44} 3369\\[-1mm]
		Hach III E~E and  Gerry C~C 1992, J. Mod. Opt. {\bf 39} 2501.
\bibitem{TCT} Epifanio G, Todorov T~C and Trapani S 1996, J. Math.
		Phys. {\bf 37} 1148.
\bibitem{Na} Naimark M~A 1976. {\it Theory of Group representations}
		( "Nauka", Moskva,\\[-1mm] 1976).
\bibitem{Br96} Brif C 1996, Ann. of Phys. {\bf 251} 180\\[-1mm]
		-- 1997  E-print quant-ph/9701003.
\bibitem{Tr96} Trifonov D~A 1996, E-prints quant-ph/9609001, quant-ph/9609017.
\bibitem{Stegun} {\it Handbook of mathematical  functions},  edited  by M.
		Abramowitz and I A  Stegun\\[-1mm]
        ({\i National bureau of standards},  1964) (Russian translation,
		M.  "Nauka", 1979).
\bibitem{XMa} Ma X and Rhodes W 1990, Phys. Rev. {\bf 41} 4624.
\bibitem{DoMaMa} Dodonov V~V, Man'ko O~V and Man'ko V~I 1994, Phys. Rev.
		A{\bf 50} 813\\[-1mm]
		Man'ko V~I 1996, E-print quant-ph/9601023 (also
        in NASA Conference \\[-1mm]
		Publication 3322, Greenbelt, MD, 1996) p. 115).
\bibitem{Sud2} Sudarshan E~C~G 1993, in NASA Conference Publication \\[-1mm]
		No. 3219 (NASA, Greenbelt, MD, 1993), p. 241\\[-1mm]
		Simon R, Sudarshan E~C~G and Mukunda N 1988, Phys. Rev. A{\bf 36}
		3868.
\bibitem{Olg} Man'ko Olga 1996, Preprint ICTP (Trieste) IC/96/39\\[-1mm]
		Dodonov V~V, Man'ko V~I and Nikonov D~E 1995, Phys. Rev. A{\bf 51}
		3328\\[-1mm]
		Ansari N~A and Man'ko V~I 1994, Phys. Rev. A{\bf 50} 1942.
\bibitem{Wun} A. W\"unsche 1995, Acta Phys. Slovaca {\bf 45} 413.
\bibitem{LuPe}  Luis A and Perina J 1996, Phys. Rev. A{\bf 53} 1886.
\bibitem{BrAr} Brif C and Ben-Aryeh Y 1996, Quantum Semiclass. Opt.
		{\bf 8} 1.
\bibitem{BrMa} Brif C and Mann A 1996, Phys. Rev. A{\bf 54} 4505.

\end{thebibliography}
\end{document}